\documentclass[aps,prd,nofootinbib,onecolumn,preprintnumbers,balancelastpage,latexsym,array,longbibliography,nobibnotes]{revtex4-2}

\usepackage{hyperref}
\usepackage{amsfonts}
\usepackage{mathrsfs}
\usepackage{epsfig}
\usepackage{graphicx}               
\usepackage{url}
\usepackage{hyperref}
\usepackage{float}
\usepackage{subcaption, caption}\usepackage{xargs} 
\usepackage{color}
\usepackage{todonotes}
\usepackage{amsmath, amssymb}

\newcommandx{\task}[2][1=]{\todo[size=\small,linecolor=red,backgroundcolor=red!25, bordercolor=red,#1]{#2}}
\newcommandx{\change}[2][1=]{\todo[size=\small,linecolor=blue,backgroundcolor=blue!25,bordercolor=blue,#1]{#2}}
\usepackage{graphicx}
\usepackage{epsf}

\newcommand{\be}{\begin{equation}}
\newcommand{\ee}{\end{equation}}
\newcommand{\bea}{\begin{eqnarray}}
\newcommand{\eea}{\end{eqnarray}}

\usepackage{comment}

\newcommand{\nc}{\newcommand}

\nc{\beq}{\begin{equation}}
\nc{\eeq}{\end{equation}}
\nc{\beqa}{\begin{eqnarray}}
\nc{\eeqa}{\end{eqnarray}}

\usepackage{slashed}

\newcommand{\lsim}{\!\mathrel{\hbox{\rlap{\lower.55ex \hbox{$\sim$}} \kern-.34em \raise.4ex \hbox{$<$}}}}
\newcommand{\gsim}{\!\mathrel{\hbox{\rlap{\lower.55ex \hbox{$\sim$}} \kern-.34em \raise.4ex \hbox{$>$}}}}

\def\be{\begin{equation}}
\def\ee{\end{equation}}

\newcommand{\Fref}[1]{Fig.\,\ref{#1}}
\newcommand{\Eref}[1]{Eq.\,(\ref{#1})}

\newcommand\affspc{\vspace{4pt}}

\usepackage{footmisc}
\usepackage{setspace}

\begin{document}

\preprint{DESY 23-125}

\title{Aspects of Particle Production from Bubble Dynamics \\
at a First Order Phase Transition}

\author{Bibhushan Shakya}

\affiliation{Deutsches Elektronen-Synchrotron DESY, Notkestr.\,85, 22607 Hamburg, Germany \affspc}

\begin{abstract}
First order phase transitions (FOPTs) constitute an active area of contemporary research as a promising cosmological source of observable gravitational waves. The spacetime dynamics of the background scalar field undergoing the phase transition can also directly produce quanta of particles that couple to the scalar, which has not been studied as extensively in the literature. This paper provides the first careful examination of various aspects of this phenomenon, which is important for understanding the dynamics of the phase transition, the generation of gravitational waves, and various high energy and beyond the Standard Model phenomena. In particular, the contributions from various stages of FOPTs (bubble nucleation, expansion, collision, post-collision) are disentangled, and conceptual aspects of the associated underlying physics relevant for particle production are clarified. Subtleties related to non-universality of particle interactions and masses in different vacua are discussed, and the suppression of nonperturbative effects such as tachyonic instability and parametric resonance due to the inhomogeneous nature of the process is examined. 

\end{abstract}

\maketitle

\tableofcontents

\newpage

\section{Introduction}

The physics of a first order phase transition (FOPT) -- the decay of the false (metastable) vacuum of a theory into the energetically favored true (stable) vacuum through the nucleation, expansion, and percolation of true vacuum bubbles \cite{Hogan:1983ixn,Witten:1984rs,Hogan:1986qda,Kosowsky:1991ua,Kosowsky:1992rz,Kosowsky:1992vn,Kamionkowski:1993fg} -- has been extensively studied in the literature for several decades, previously in the context of inflation \cite{Guth1981,LaSteinhardt1989}, and more recently as a promising cosmological source of gravitational waves (GWs)\cite{Grojean:2006bp,Caprini:2015zlo,Caprini:2018mtu,Caprini:2019egz,Athron:2023xlk} that could be detected with upcoming GW experiments. While the Standard Model (SM) does not admit any FOPTs, such transitions can be readily realized in extended sectors in many realistic beyond the Standard Model (BSM) scenarios \cite{Schwaller:2015tja,Jaeckel:2016jlh,Dev:2016feu,Baldes:2017rcu,Tsumura:2017knk,Okada:2018xdh,Croon:2018erz,Baldes:2018emh,Prokopec:2018tnq,Bai:2018dxf,Breitbach:2018ddu, Fairbairn:2019xog, Helmboldt:2019pan,Ertas:2021xeh,Jinno:2022fom}. The dynamics of the bubble walls separating the true and false vacua is known to produce detectable GWs through various means: through the scalar field energy densities in the bubble walls after collision\,\cite{Kosowsky:1991ua,Kosowsky:1992rz,Kosowsky:1992vn,Kamionkowski:1993fg,Huber:2008hg,Bodeker:2009qy,Jinno:2016vai,Jinno:2017fby,Konstandin:2017sat,Cutting:2018tjt,Cutting:2020nla}, the production of sound waves (SWs)~\cite{Hindmarsh:2013xza,Hindmarsh:2015qta,Hindmarsh:2017gnf,Cutting:2019zws,Hindmarsh:2016lnk,Hindmarsh:2019phv} and turbulence~\cite{Kamionkowski:1993fg,Caprini:2009yp,Brandenburg:2017neh,Cutting:2019zws,RoperPol:2019wvy,Dahl:2021wyk,Auclair:2022jod} in the plasma in the presence of significant interactions, or through energy transfer to nontrivial spatial configurations of feebly-interacting particles \cite{Jinno:2022fom}. 

In addition to GWs, such phase transitions can also be responsible for particle production: any particle that couples to the background scalar field undergoing the phase transition gets produced due to the spacetime dynamics of the scalar field over the duration of the transition. Note that particle production should be a significantly stronger phenomenon than GW emission since the graviton coupling is Planck suppressed, whereas particle couplings tend to be significantly larger. Furthermore, while GWs can only be produced after the bubbles collide as they correspond to transverse, traceless excitations of the metric, which cannot be sourced by spherically symmetric bubbles, particle production is more general and can occur over all stages of FOPTs, including the nucleation and expansion of spherical bubbles. In the GW literature, particle production from FOPTs is primarily considered in the context of interactions between the fast moving bubble walls and the surrounding plasma \cite{Bodeker:2017cim,Hoche:2020ysm,Azatov:2020ufh,Gouttenoire:2021kjv,Baldes:2023fsp,Ai:2023suz,Long:2024sqg}. While such plasma-induced particle production effects can be significant, in this paper we concern ourselves with the more fundamental, irreducible form of particle production that occurs purely due to the dynamics of the background field itself. Particle production from a changing background field is a well-known physical phenomenon familiar from various contexts, such as gravitational particle production \cite{Parker:1969au,Grib:1969ruc,Zeldovich:1971mw,Kolb:2023ydq}, Schwinger effect \cite{Schwinger:1951nm}, or Hawking radiation from black holes \cite{Hawking:1974rv,Hawking:1975vcx}. Such particle production, which exists even in the absence of a plasma, could be particularly relevant for phase transitions where a plasma is absent (such as phase transitions in vacuum) or diluted (such as supercooled transitions (see e.g.\,\cite{Randall:2006py,Konstandin:2011dr,vonHarling:2017yew,Ellis:2018mja,Baratella:2018pxi,Bruggisser:2018mrt,DelleRose:2019pgi,VonHarling:2019rgb,Fujikura:2019oyi,Ellis:2019oqb,Brdar:2019qut,Baldes:2020kam})). In cases where a thermal plasma is present, the details of particle production can still be important for deducing the GW spectra as well as for various non-thermal particle production applications to BSM physics, in particular since the bubble walls can be ultrarelativistic, producing particles with masses far heavier than the scale of the phase transition or the temperature of the thermal bath (as has been explored, e.g.\,in  \cite{Falkowski:2012fb,Katz:2016adq,Freese:2023fcr,Giudice:2024tcp,Cataldi:2024pgt}). 

The calculation of particle production from background field dynamics during various stages of a FOPT is complicated due to the inhomogeneous nature of the process. Standard particle production calculations performed in homogeneous time-varying backgrounds are inapplicable, necessitating more complex semi-analytic and numerical approaches to the problem. At the same time, numerical studies targeted towards computing GW spectra yield little insight into the particle production process, since such studies focus on macroscopic (Hubble scale) setups and generally cannot resolve the particle physics scales relevant for particle production. Only a handful of papers in the literature have addressed this important phenomenon: the formalism to study this process was first presented in \cite{Watkins:1991zt}, followed by \cite{Konstandin:2011ds}, and further developed in \cite{Falkowski:2012fb} with semi-analytic results for some idealized cases, and most recently with numerical studies for more realistic cases in \cite{Mansour:2023fwj}, and in \cite{Giudice:2024tcp} with a discussion of the gauge dependence of the formalism.

The purpose of this paper is to clarify and expand on these previous works \cite{Watkins:1991zt,Konstandin:2011ds,Falkowski:2012fb,Mansour:2023fwj} and provide a more comprehensive understanding and discussion of particle production from FOPTs. While these papers provide the tools necessary to calculate particle production using the Fourier transform of the field's spacetime configuration, conceptual understanding of various physical aspects of the process is lacking; this paper clarifies several such aspects.  In particular, we discuss the contributions from the various stages of the FOPT (nucleation, expansion, collision, and post-collision). The previous papers \cite{Watkins:1991zt,Konstandin:2011ds,Falkowski:2012fb} only considered particle production from the collision and post-collision phases, assuming the contribution from the nucleation and expansion phases to be subdominant. This paper provides the first qualitative estimates of particle production from bubble nucleation and expansion, verifying that these stages indeed provide subdominant contributions. A deeper understanding of the underlying physics of particle production from bubble collisions is then provided through examinations of some illuminating simple and realistic cases. We will also discuss non-perturbative, resonant particle production processes, such as tachyonic instability and parametric resonance, finding that such processes are suppressed by the inhomogeneous nature of FOPTs. 

The paper is structured as follows. Section \ref{sec:nucleation} reviews the standard approach to calculate particle production using Bogoliubov transformations, and applies it to the cases of bubble nucleation.  Section \ref{sec:expand} considers the bubble expansion phase, examining particle production from constant velocity and accelerating bubble walls.  Section \ref{sec:expcol} discusses the formalism for calculating particle production from the collision and post-collision phases, and studies some illuminating simple cases as well as more realistic configurations to provide a clearer understanding of the phenomenon. Section \ref{sec:particlephysics} discusses subtleties related to the non-universality of particle interactions and masses in the two vacua, as well as the (ir)relevance of non-perturbative effects in such inhomogeneous processes. The main results of the paper are summarized in Section \ref{sec:summary}.

\section{Bubble Nucleation}

\label{sec:nucleation}

A first order phase transition can be considered in three stages: (i) it begins with the nucleation of bubbles of true vacuum (broken phase) in a background of false vacuum (unbroken phase); (ii) most of the space is converted into the broken phase via the expansion of these bubbles; (iii) the transition completes when these bubbles collide with each other and percolate. Each stage features distinct inhomogeneous forms of scalar field evolution, which can give rise to particle production. 

In this section we first outline the standard particle production calculation in a homogeneous transition, where particle production due to the changing background field can be calculated using the well-known Bogoliubov transformation between mode expansions of particles in the two (unbroken and broken) phases. We then apply this formalism to the stage of bubble nucleation, which can be treated as a homogeneous process in the thin-wall approximation.

\subsection{Homogeneous Case: Bogoliubov Transformation}

Details of the standard Bogoliubov transformation calculation can be found in several textbooks \cite{Rubakov:2017xzr,Birrell:1982ix,Parker:2009uva} and papers (e.g.\,\cite{GarciaBellido:2001cb,Tanaka:1993ez,Yamamoto:1994te,Hamazaki:1995dy,Mohazzab:1995zx,Maziashvili:2003kj,Maziashvili:2003sk,Giudice:2019iwl}), and we will not repeat them here, but only cover aspects that will be relevant for our subsequent calculations.    

Consider a homogeneous (second order) phase transition, where the vacuum expectation value (vev) of a scalar field $\phi$ changes coherently over all space, increasing from $0$ to $v_{\phi}$ over a timescale $\mathcal{T}$ as
\be
\langle \phi \rangle (t) = v_{\phi}(t)=\frac{1}{2}v_{\phi} [1+\tanh (t/\mathcal{T}) ].
\label{eq:tanh}
\ee
Consider a scalar $\psi$ that obtains its mass from the scalar vev, $m_\psi^2 (t) = \lambda^2 (v_{\phi} (t) )^2$, where $\lambda$ is a dimensionless coupling constant. The Klein Gordon (KG) equation for $\psi$ (ignoring all other interactions for simplicity) is
\be
(\partial_t^2-\partial_x^2+m_\psi^2 (t))\,\psi(t,x)=0\,.
\label{kge1}
\ee
The $\psi$ field can be decomposed into momentum modes $\psi_k$ that satisfy the mode equation 
\be
(\partial_t^2+k^2+m_\psi^2 (t))\,\psi_k (t)=0\,.
\label{eq:modes}
\ee
The essence of the calculation of particle production from vacuum (for details, see   \cite{Rubakov:2017xzr,Birrell:1982ix,Parker:2009uva}) is as follows:
The modes can be written in terms of  positive and negative frequency oscillators $e^{i\omega t},\,e^{-i\omega t}$, where $\omega^2=k^2+m_\psi^2 (t)$; the vacuum state of the field corresponds to the configuration where all negative frequency modes are occupied, but all positive modes (particle excitations) are empty. The frequency $\omega^2 (t)=p^2+m_\psi^2(t)$ changes during the phase transition due to the change in the background field, causing $\textit{mixing}$ between negative and positive frequency modes, leading to the emergence of positive frequency modes in the final state (defined with respect to the true vacuum) even when the initial state (defined with respect to the false vacuum) is composed purely of negative frequency modes, which is interpreted as particle production.

The Bogoliubov transformation calculation assumes that the background change is sudden (i.e.\,non-adiabatic), otherwise the field simply tracks the change in the background field smoothly and remains in the vacuum state, so that no particle production occurs. The condition for non-adiabaticity is $|\omega'/\omega^2| \geq 1$. Since the mass changes by $m$ over a timescale $\mathcal{T}$, we have $\omega'\sim m/\mathcal{T}$, whereas $\omega^2\approx m^2$. Thus the evolution is non-adiabatic only for $m\mathcal{T}\leq 1$, and the results below are only valid in this regime. 

For the above ansatz (Eq.\,\ref{eq:tanh}), the mode occupation numbers for the produced particles can be expressed in terms of analytic solutions. The mode occupation number for bosons with momentum $k$ is calculated to be  \cite{Herranen:2015ima}
 \be
n^B_k =|\beta_k|^2=\frac{\sinh^2(\frac{1}{2}\pi \,\mathcal{T}\,(\omega_2-\omega_1))}{\sinh(\pi\,\mathcal{T}\,\omega_1)\sinh(\pi \,\mathcal{T}\,\omega_2)}\,, 
\label{ndensityb}
\ee
with $\omega_1\equiv k$ and $\omega_2\equiv \sqrt{k^2+m_\psi^2}$ (where $m_\psi$ is the final/asymptotic mass in the true vacuum), and $\beta_k$ is the Bogoliubov coefficient of the positive frequency $k$ mode in the final state. The mode occupation number for fermions can be calculated analogously as \cite{GarciaBellido:2001cb}:
\be
n^F_k =\frac{\cosh(\pi \,\mathcal{T}\, m_F)-\cosh(\pi \,\mathcal{T}\,(\omega_2-\omega_1))}{2\sinh(\pi\,\mathcal{T}\,\omega_1)\sinh(\pi \,\mathcal{T}\,\omega_2)}\, , 
\label{ndensityf}
\ee
where $m_F$ is the mass of the fermion after the phase transition.

The particle number density of a field $X$ is obtained by integrating the above occupation numbers over all momenta:
\be
n^X =\frac{g_X}{2\pi^2}\, \int^\infty_0 dk \, k^2\, n_k^X\,, 
 \label{particlenumber}
\ee
where $g_X$ is the number of degrees of freedom in field $X$. 

Note that in the limit $m_X\to 0$, we get $n_k^X\to 0$ for bosons as well as fermions: as the field equation of motion remains unchanged in both vacua, the field is not sensitive to the changing background, and no particle production takes place. We also note two limiting behaviors of the above expressions: 
(i) they reduce to an exponential suppression in the large $k$ limit $k \mathcal{T}\gg 1$: number densities for momenta larger than the relevant energy scale of the process, $\mathcal{T}^{-1}$, are exponentially suppressed; and (ii) in the low momentum limit $k \mathcal{T}\ll 1$, $n_k$ becomes momentum-independent and approaches a constant value. 

Integrating Eq.\,\ref{particlenumber} over all momenta gives the following particle number density:
\be
n_X =\frac{g_X}{2 \pi^2 (2\pi)^3} \,\mathcal{T}^{-3}\,\mathcal{I}\,(\mathcal{T}\,m_X)\,. 
 \label{particlenumbersimpler}
\ee
As could have been anticipated from dimensional analysis, the number density of particles produced scales as $\mathcal{T}^{-3}$, which is the relevant physical scale in the problem.  The dimensionless integral factor $\mathcal{I}(\mathcal{T}\,m_X)$ is 
\bea
\mathcal{I}(a)&\equiv&  \int^\infty_0 dx\, x^2\, \times \frac{\sinh^2\Big[\frac{1}{4} \Big(\sqrt{a^2+x^2}-x\Big)\Big]}{\sinh\Big(\frac{1}{2}\sqrt{a^2+x^2}\Big)\sinh\Big(\frac{1}{2} x\Big)}~~~~~~~(\text{bosons)}\nonumber\\
\mathcal{I}(a)&\equiv&  \int^\infty_0 dx\, x^2\, \times \frac{\cosh\Big(\frac{1}{2}a\Big)-\cosh\Big[\frac{1}{2} \Big(\sqrt{a^2+x^2}-x\Big)\Big]}{2\sinh\Big(\frac{1}{2}\sqrt{a^2+x^2}\Big)\sinh\Big(\frac{1}{2} x\Big)}~~~~~~~(\text{fermions)}
\label{eq:integralfactors}
\eea
and encapsulates the efficiency of the changing background for producing particles with various masses. For $ m_X \mathcal{T} < 1$, one finds $\mathcal{I}(a) \approx a^2$; as expected, the integral vanishes in the limit $m_X\to 0$. For $m_X \mathcal{T}\gtrsim 1,~\mathcal{I}(a)\approx 1$. In all cases, the largest contribution to the number density comes from momenta $k \mathcal{T}\sim 1$, hence particles are primarily produced with momenta close to the inverse scale of the transition, $k\sim \mathcal{T}^{-1}$.

It is important to keep in mind that these formulae are only applicable if the relevant particle gets its mass from the background field. For fields with bare masses much larger than that contributed by the background field, the production gets exponentially suppressed; this occurs when the non-adiabaticity condition is violated, i.e. when $\delta m / m \ll m \mathcal{T} $.

\subsection{Particle Production at Bubble Nucleation}
\label{subsec:nucleation}

We can use the above results for the homogeneous case to estimate particle production from the process of bubble nucleation.  A first order phase transition commences with bubbles of true vacuum nucleating spontaneously with a critical radius $R_c$. The nucleation rate as well as field profile inside the nucleated bubble are determined by the bounce action. We will work in the thin wall approximation (applicable when the difference in energy between the true and false minima is smaller than the height of the barrier separating the two), for which the field inside the bubble is in the true vacuum immediately following nucleation, separated from the false vacuum region outside by a wall of thickness $l_{\mathrm{w}}\!\ll \!R_c$  \footnote{For thick-walled bubbles, additional scalar field dynamics inside the bubble can lead to additional particle production (see e.g.\,\cite{Cutting:2020nla}). We ignore such possibilities in this paper since this contribution is expected to be subdominant to the leading contribution from bubble collisions.}. The background field configuration within the thin wall is given by the ansatz
\be
v_{\phi}(x)=\frac{1}{2} v_{\phi}\,[1+ \tanh (x/l_{\mathrm{w}})]  \label{wallprofile}
\ee
for the wall centred at $x=0$. This ansatz is exact for a quartic potential but remains applicable to the extent that a given potential can locally be approximated as a quartic \cite{Kolb:1990vq, Mazumdar:2018dfl}. Note the analogy with the ansatz used for the homogeneous transition (Eq.\,\ref{eq:tanh}) in the previous subsection.

When the thin wall approximation is justified, we can consider the scalar field within the bubble to be homogeneous. Although bubbles of critical radii are assumed to nucleate instantaneously, the evolution of the scalar field from the false to true vacuum inside the nucleated bubble is also determined by the dynamics of the bounce action. Since the bounce action is O(4) symmetric, we can posit that the temporal evolution (of the nucleating bubble) also follows the same ansatz as its spatial variation, i.e. inside the nucleated bubble we posit that the field evolves as
\be
v_{\phi}(t)=\frac{1}{2} v_{\phi}\,[1+ \tanh (t/l_{\mathrm{w}})] ~~~~~ \text{for } r<R_c\,.  \label{VEVt}
\ee
Therefore, the wall thickness $l_{\mathrm{w}}$ also parameterizes the timescale of field evolution at bubble nucleation.

 Thus, the number density of particles produced during bubble nucleation can be estimated from the homogeneous case Eq.\,\ref{particlenumbersimpler}, \ref{eq:integralfactors} with the replacement $\mathcal{T}\to l_w$.\,\footnote{These simple estimates are in qualitative agreement with a more rigorous but complex calculation in \cite{Yamamoto:1994te}.} This homogeneous approximation should be valid for modes $k\gg R_c^{-1}$. In principle, this implies that Eq.\,\ref{eq:integralfactors} should be calculated with an IR cutoff $2\pi l_{\mathrm{w}}/R_c$; however, in practice, since the integrals over phase space are known to be dominated by $k\sim l_{\mathrm{w}}^{-1}$, and $l_{\mathrm{w}}\ll R_c$ in the thin wall approximation, the implementation of this IR cutoff is not necessary. 
 
 It is important to keep in mind that the above particle production only takes place at sites of bubble nucleation, i.e.\,within spheres of radii $R_c$. These number densities will eventually get diluted by a factor $(R_c/R_*)^3$ when the particles diffuse out to fill the entire volume of the bubble at its maximal size at collision, corresponding to radius $R_*$. Accounting for this, the final number density of particles from bubble nucleation in the thin wall approximation is given by  
\be
n_X \approx\frac{g_X}{4 \pi^5} \,l_w^{-3}\,\left(\frac{R_c}{R_*}\right)^3\mathcal{I}\,(l_w\,m_X)\,, 
 \label{nucleationcontribution}
\ee
with $\mathcal{I}(a)$ given by \Eref{eq:integralfactors}. Again, this formula, derived from a Bogoliubov tranformation calculation, is only valid when the evolution is non-adiabatic, which occurs only for $l_w\,m_X \leq 1$. 

Due to this significant volume dilution factor, the contribution from bubble nucleation is generally negligible compared to the other stages of the phase transition. However, it is worth noting that if the energy density in particles as calculated from Eq.\,\ref{nucleationcontribution} becomes comparable to the vacuum energy released in the nucleation process, the backreaction from particle production should be appropriately taken into account in calculating the nucleation rate of critical bubbles.  

\section{Bubble Expansion}
\label{sec:expand}

Next, let us consider the stage of bubble expansion, where the bubble walls propagate out in space. This stage is responsible for converting most of the space from the false vacuum to the true vacuum. The expansion of bubbles also involves an evolution of the background field from the false to true vacuum, $0\to v_{\phi}$; one might therefore expect similar particle production as in the homogeneous case or bubble nucleation as a consequence of this dynamics. In fact, a naive implementation of the above formalism would suggest that particle number densities in this case scale as $\sim(\gamma_w/l_w)^3$ (where $\gamma_w$ is the Lorentz boost factor of the bubble wall), which is the timescale over which a point in space transitions from the false to true vacuum as the wall passes through. However, the transition in this case is $\textit{inhomogeneous}$; the background field evolves in space as well as time, which complicates the calculation and, as we will see below, changes the result completely. There are two distinct scenarios that need to be considered: bubble walls propagating through space  with constant velocity, which occurs if frictional forces from the plasma cause the bubble walls to reach terminal velocity, and accelerating bubble walls, which occurs if frictional forces are weaker than the outward pressure caused by the latent vacuum energy released in the transition, causing the bubble walls to achieve the so-called runaway behavior. 
We will consider both cases in turn. 

\subsection{Propagating Bubble Wall at Constant Velocity}
\label{subsec:single}

We will consider a single, planar bubble wall propagating at constant velocity, for which it is possible to tackle the problem analytically using the Bogoliubov transformation approach. For simplicity, consider a bubble wall of constant thickness $l_w$ and profile given in Eq.\,\ref{wallprofile} sweeping through space with velocity $v_{\mathrm{w}}$. This simplification is justified for particle production considerations since particle production is a local process that should not be affected by the extended shape and dynamics of the entire bubble\,\footnote{Throughout this paper we assume that the bubbles are perfectly spherically symmetric; for studies of asymmetric fluctuations and inhomogeneities on bubble walls, see \cite{Braden:2014cra,Braden:2015vza,Bond:2015zfa}.}. For this configuration, the KG equation from \Eref{kge1} becomes
\be
\left(-\frac{\partial^2}{\partial t^2}+\frac{\partial^2}{\partial x^2}\right)\psi(t,x)=\frac{m_\psi^2}{4}\left(1-\text{tanh}\left(\frac{v_w t-x}{l_{\mathrm{w}}}\right)\right)^2\psi(t,x)\,.
\label{kgew}
\ee

Since the mass term now depends on both $t$ and $x$, it is not possible to separate this equation into independent mode equations in the $(x,t)$ basis as in the homogeneous case (Eq.\,\ref{eq:modes}). One can nevertheless perform a change of coordinates that allows for such decomposition, enabling us to analytically compute particle production using the standard Bogoliubov transformation approach as above: start with an initial state where all negative frequency modes are occupied, decompose the KG equation mode-by-mode in the appropriate basis, compute the mixing of modes induced by the phase transition, and look at nonzero coefficients of positive frequency modes in the final state as a signal of particle production across the process. This calculation is presented in detail in Appendix \ref{appendixsinglewall}, and gives the following result: the condition for particle production out of vacuum is
\be
p_2^2(v_w^4-1)>m_\psi^2(1-v_w^2)\,.
\ee
Here $p_2$ is a conjugate momentum in the new coordinates (see Appendix \ref{appendixsinglewall} for details). 
We see that this condition cannot be satisfied for any real values of $p_2,\, m_\psi$ if $v_w<1$. This leads us to the intriguing conclusion that a single bubble wall that propagates with a constant velocity below the speed of light cannot excite particles (positive frequency modes) starting with an initial vacuum state (negative frequency modes).  While it is instructive to go through the proper calculation (Appendix \ref{appendixsinglewall}), this result should also be obvious from intuition: for a bubble wall traveling at constant speed $v_w<1$, one can always perform a Lorentz boost to the rest frame of the wall; in this frame, the configuration is perfectly static, hence no particle excitations can occur. 

\subsection{Accelerating Bubble Wall}
\label{subsec:accelerating}

As the latent energy released from the false vacuum is transferred to the bubble walls, if opposing frictional forces from the plasma are not strong enough to counteract this outward pressure, the bubble walls continuously accelerate rather than propagate at constant velocity. Accelerating objects emit radiation, hence some particle production is expected to occur from accelerating bubble walls. This contribution has not been calculated precisely in the literature, and it is unclear how to perform this calculation. However, we can make some simple heuristic parametric estimates here, which suggest that the energy transferred to the produced particles is small compared to the vacuum energy density, which will be sufficient for the purposes of this paper. 

We will again work in the rest frame of the bubble wall. This frame is not an inertial frame; however, it allows us to make some qualitative arguments by making use of the equivalence principle. According to the equivalence principle, a non-uniformly accelerating bubble wall can be considered equivalent to a bubble wall at rest in a changing gravitational field. We can therefore estimate particle production by making use of the familiar phenomenon of gravitational particle production in a changing background gravitational field (see e.g.\,\cite{Kolb:2023ydq} for a recent review), which is known to yield a particle number density of order $H^3$, where $H$ is the Hubble parameter. 

To calculate the instantaneous acceleration of the bubble wall, we can make use of the Lorentz boost factor of the bubble wall, $\gamma_w=(1-v_w)^{-1/2}$, which is known to grow linearly with the size of the bubble, $\gamma_w\approx R/R_c$ (see e.g.\,\cite{Ellis:2020nnr}) in the absence of strong frictional forces, where $R_c$ is the critical radius of the bubble at nucleation. Taking the time derivate of this relation, we obtain the wall acceleration $a_w\approx R_c^2/R^3$, as well as the following relations between the parameters, assuming relativistic bubbles ($v_w\approx 1$):
\beq
-\frac{\dot{a}_w}{a_w}\approx\frac{3 \dot{\gamma}_w}{\gamma}\approx\frac{3 \dot{R}}{R}\approx\frac{3}{R}\,.
\eeq
We see that the inverse size of the bubble $R^{-1}$ parametrizes the rate at which the physical quantities of the setup are changing, playing the role analogous to the Hubble rate for gravitational particle production. Therefore, it is reasonable to posit that the instantaneous rate of particle production, analogous to gravitational particle production, should scale as $\sim R^{-3}$. Since only fields that couple to the background field undergoing the phase transition can get produced through the dynamics of the wall, we further assume that the rate must be proportional to $g_X^2\,R^{-3}$, where $g_X$ is the coupling between the produced particle $X$ and the background field. Integrating this quantity over the motion of the bubble wall of a spherical bubble, we can estimate that the average number density of particles $X$ produced from the wall motion from production to collision is
\beq
n_X\approx \frac{g_X^2 \int^{R_*}_{R_c}  R^{-3}\cdot 4\pi R^2\cdot dR}{\frac{4}{3}\pi R_*^3} \approx g_X^2 R_*^{-3}\, \text{ ln}(R_*/R_c)\,,
\eeq
where $R_*$ is the maximal size of the bubble at collision, and in the second expression we have dropped $\mathcal{O}(1)$ factors. For realistic phase transition parameters, $R_*\sim H^{-1}$, whereas $R_c\sim v_\phi^{-1}$,  thus the logarithmic factor is $\mathcal{O}(1)$ - $\mathcal{O}(10)$. Therefore, we estimate that the number density of particles produced due to bubble acceleration over the expansion phase of the bubbles is parametrically $\sim g_X^2\,R_*^{-3}\sim g_X^2\,H^3$. Crucially, note that this is significantly smaller than a thermal number density $\sim T^3$; consequently, the energy density in the produced particles is generally significantly smaller than the energy released in the phase transition,  and particle production is therefore unlikely to affect bubble dynamics during the expansion phase. It will also turn out to be subdominant to the energy transferred to particles during bubble collisions, discussed in the next section.

\section{Bubble Collision and Post-Collision Dynamics}
\label{sec:expcol}

The bubble expansion stage is followed by the collision of bubbles, resulting in local excitations of the background field that create scalar waves that propagate out from the collision points. This stage is far too complicated to be solved with analytic or simple heuristic approaches from the previous sections \footnote{For some studies of kink-antikink scattering, which holds some similarity to the bubble collision configurations considered here, see \cite{Mukhopadhyay:2021wmu,Mukhopadhyay:2023zmc}.}.  Instead, a modified numerical approach -- analogous to the approach used to compute gravitational wave emission from such configurations -- becomes necessary to solve such systems, which will be the topic of this section. 

Depending on the details of the scalar potential, the collisions can be elastic or inelastic (see discussions in \cite{Watkins:1991zt,Konstandin:2011ds, Falkowski:2012fb, Jinno:2019bxw, Mansour:2023fwj}):

\textbullet\ \textit{Elastic collisions} correspond to two colliding bubble walls reflecting off each other, re-establishing the false vacuum in the region in between. This is followed by multiple repeated collisions and reflections until the true vacuum is eventually realized. 

\textbullet\ \textit{Inelastic collisions} refer to the realization of true vacuum upon bubble collision, with the energy in the bubble walls getting converted to scalar field oscillations. 
 
In both cases, the scalar field at the point of collision gets excited to a field value away from the minima, resulting in oscillations around the corresponding minimum: true (false) minimum for inelastic (elastic) collisions. Such dynamics of the background field can give rise to particle production.

 \subsection{Formalism}
 \label{sec:formalism}

This subsection briefly summarizes the formalism for calculating particle production from FOPTs, as introduced in \cite{Watkins:1991zt} and further developed in \cite{Konstandin:2011ds, Falkowski:2012fb, Mansour:2023fwj}. The approach consists of evaluating the imaginary part of the effective action of the background field, which essentially amounts to taking the Fourier transform of the classical external field configurations $\phi (x,t)$ to decompose the scalar field dynamics into modes of definite four-momenta $\chi=\omega^2-k^2 >0$, which are interpreted as off-shell propagating field quanta of $\phi$ with mass $m^2=\chi$. Each mode excitation can decay, as determined by the imaginary part of its 2-point 1 particle irreducible(1PI) Green function. The number of particles produced per unit area of bubble wall (assuming decay into a pair or identical particles) can then be calculated as \cite{Watkins:1991zt,Konstandin:2011ds,Falkowski:2012fb}
\be
\frac{N}{A}= 2 \int\frac{dk\,d\omega}{(2\pi)^2}\,|\tilde{\phi}(k,\omega)|^2 \,\text{Im}[\tilde{\Gamma}^{(2)}(\omega^2-k^2)]\,.
\label{particle}
\ee
Here, $\tilde{\phi}(k,\omega)$ is the Fourier transform of the field configuration $\phi(x,t)$, and the imaginary part of the 2-point 1PI Green function is given by 
\be
\text{Im} [\tilde{\Gamma}^{(2)}(\chi)]=\frac{1}{2}\sum_\alpha \int d\Pi_\alpha |{\mathcal{M}}(\phi\to\alpha)|^2\,\Theta(\chi-\chi_{min\,(\alpha)}).
\label{optical}
\ee
Here $\alpha$ sums over all final particle states that can be produced, $|{\mathcal{M}}(\phi\to\alpha)|^2$ is the spin-averaged squared amplitude for the corresponding decay channel, integrated over the relativistically invariant n-body phase space $d\Pi_\alpha$, and $\Theta$ is the Heaviside step function restricting the integral to the kinematically allowed region $\chi>\chi_{min}=(\sum m_\alpha)^2$. Gauge-dependence issues related to evaluating the decay probability of such off-shell excitations are discussed in \cite{Giudice:2024tcp}. Changing to variables $\chi=\omega^2-k^2$ and $\xi=\omega^2+k^2$, the above expression can be simplified to \cite{Falkowski:2012fb}
\be
\frac{N}{A}=\frac{1}{4\pi^2}\int_{\chi_{min}}^{\chi_{max}} d\chi\,f(\chi) \,\text{Im} [\tilde{\Gamma}^{(2)}(\chi)]\,,
\label{number}
\ee
with the integral over $\xi$ absorbed into $f(\chi)$. 
Here the efficiency factor $f(\chi)$ encapsulates the spectrum of excitations of the background field. The upper cutoff $\chi_{max}=(2 \gamma_w/l_w)^2$ denotes the maximum energy scale present in the system, given by the inverse Lorentz boosted bubble wall thickness. The lower limit $\chi_{min}$ is determined by the bubble size $R_*$ at collision. Thus $f(\chi)$ encodes information about the spacetime configuration and dynamics of the background field, whereas $[\tilde{\Gamma}^{(2)}(\chi)]$ encodes the particle physics aspects of its decay. 
 
Here, note that $N/A$ in \Eref{number} represents the produced particle number per unit surface area of bubble walls. The particles will eventually diffuse uniformly throughout the volume of the bubbles, so that the final particle number
density is given by \cite{Falkowski:2012fb}
\be
n=\frac{N}{A}\,\frac{2}{3\,R_*}=\frac{1}{6\pi^2\,R_*}\int_{\chi_{min}}^{\chi_{max}} d\chi\,f(\chi) \,\text{Im} [\tilde{\Gamma}^{(2)}(\chi)]\,,
\label{eq:ndensity}
\ee
where $R_*$ is the maximal radius of the bubble (at collision).

\subsection{Simple Cases}
\label{sec:simple}

\begin{figure}
     \begin{subfigure}{0.31 \textwidth}
     \includegraphics[width=\textwidth]{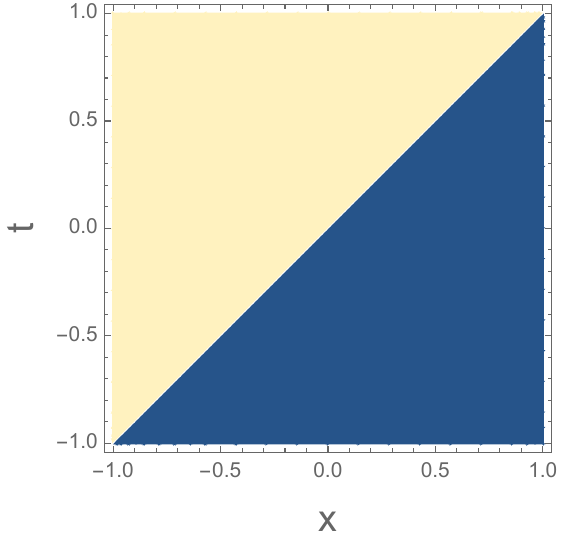}
     \caption{Single bubble wall propagating at constant velocity} 
       \end{subfigure}  
      \begin{subfigure}{0.32 \textwidth}
     \includegraphics[width=\textwidth]{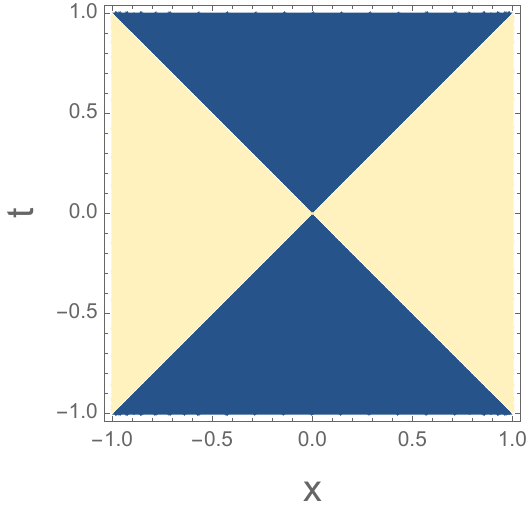}
     \caption{Two bubble walls colliding at the origin}  
     \end{subfigure}
     \begin{subfigure}{0.32 \textwidth}
     \includegraphics[width=\textwidth]{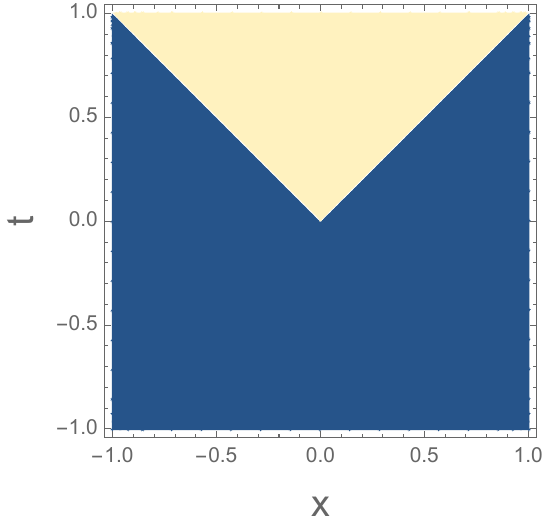}
     \caption{Single bubble expanding out from the origin}  
     \end{subfigure}
     \caption{Spacetime evolution of some simple field configurations of planar thin-walled bubbles, plotted for wall speed $v_w\approx 1$, with $t,x$ in units of $v_\phi^{-1}$. The yellow (light) and blue (dark) regions corresponds to the true and false vacua respectively.}
     \label{fig:simpleconfigurations}
 \end{figure}

We now study the Fourier transform of some simple configurations, shown in Figure \ref{fig:simpleconfigurations}, to clarify several important aspects of the underlying physics that will be relevant for more realistic cases. For simplicity, consider two dimensional (1+1) systems, with the false and true vacua at $0$ and $v_{\phi}$ respectively, separated by infinitesimally thin walls (step functions). For these simplified setups, the Fourier transforms can be calculated analytically. 

\vskip 0.2cm
\noindent \textit{(a) Single bubble wall moving at constant velocity:}

First, consider a single bubble wall moving at constant velocity $v_{\mathrm{w}}$, corresponding to the configuration studied above in Sec.\,\ref{subsec:single}. The field configuration is given by (see Fig.\,\ref{fig:simpleconfigurations}(a)\,)
\be
\phi(x,t)=v_{\phi}\, \Theta( v_{\mathrm{w}} t - x)\, 
\ee
where $\Theta$ is the Heaviside step function. Its Fourier transform is 
\be
\tilde{\phi}(k,\omega)=
-2\pi i \frac{v_{\phi} \delta(k\, v_{\mathrm{w}} + \omega)}{k}
 + 2\pi^2\, v_{\phi}\, \delta(k)\,\delta(k\,v_{\mathrm{w}} + \omega)\,,
\ee
where $\delta$ is the Dirac delta function. Note that the second term vanishes for all nonzero $\omega, k$. The Dirac delta function in the first term forces $\omega=-k\,v_{\mathrm{w}}$, hence $\omega^2< k^2$ for $v_{\mathrm{w}}<1$, and no $\chi=\omega^2-k^2>0$ modes are excited in this case, implying no particle production occurs according to the formalism in the previous subsection. This is consistent with the result in Sec.\,\ref{subsec:single} that no particle production occurs from a single wall moving at constant velocity $v_{\mathrm{w}}<1$. It is intriguing to note that $v_{\mathrm{w}}>1$ does allow for particle production, also consistent with the observations in Sec.\,\ref{subsec:single}.

\vskip 0.2cm
\noindent \textit{(b) Two bubble walls in a perfectly elastic collision:}

Next, consider the case where two planar bubble walls, each with speed $v_{\mathrm{w}}$, approach each other and undergo a perfectly elastic collision, getting reflected with the same speeds $v_w$, re-establishing the false vacuum in the region in between (see Fig.\,\ref{fig:simpleconfigurations}(b), with the collision occurring at the origin). This corresponds to the perfectly elastic collision configuration studied in \cite{Watkins:1991zt,Konstandin:2011ds, Falkowski:2012fb}. The field configuration in this case can be written as
\be
\phi(x,t)= 
\begin{cases}
 0 \text{~~~~~~for~} t^2>x^2\\
 v_{\phi}\text{~~~~~~for~} t^2\leq x^2
\end{cases}
\label{eq:cases}
\ee
The Fourier transform for such cases is known to be 
 \cite{Watkins:1991zt, Falkowski:2012fb}
\be
\tilde{\phi}(k,\omega)=\frac{4\, v_{\phi}}{\omega^2-k^2\,v_{\mathrm{w}}^2}\,.
\label{eq:FTelastic}
\ee
As we can see, this configuration does excite $\chi=\omega^2-k^2>0$ modes, resulting in particle production. A-priori, it is not clear whether particle production occurs due to the relative motion of the two bubble walls, or entirely from the collision process. In Sec.\,\ref{subsec:single} we argued that a single bubble wall at constant velocity cannot produce particles; however, it is not clear whether the argument applies to two bubbles moving relative to each other, since now there is no reference frame where the scalar field configuration is completely static. However, Ref.\,\cite{Watkins:1991zt} also considered the case of two bubble walls passing through each other without interacting (as occurs in sine-Gordon type theories), finding no particle production in such cases, indicating that particle production is entirely due to the collision between the bubble walls. 

\vskip 0.2cm
\noindent \textit{(c) Single expanding bubble:}

Next, consider a single bubble that starts as a point (at the origin) and expands outwards at constant speed $v_{\mathrm{w}}$ (see Fig.\,\ref{fig:simpleconfigurations}(c)\,). The field configuration in this case can be written as
\be
\phi(x,t)= v_{\phi} \,\Theta(v_{\mathrm{w}}\,t - |x|)\,
\label{eq:newcase}
\ee
and yields the Fourier transform  
\be
\tilde{\phi}(k,\omega)=\frac{2\, v_{\phi}}{\omega^2-k^2\,v_{\mathrm{w}}^2}\,.
\label{eq:singlebubble}
\ee  
Incredibly, this is the same as the Fourier transform for the case of an elastic bubble collision, \Eref{eq:FTelastic}, with half the amplitude (originating from the fact that the scalar field only changes in half the plane in this case). This suggests that the mode excitations in \Eref{eq:FTelastic} originate from the ``turning around" of the wall at the origin. In other words, rather than some complicated dynamics from the collision of the two walls, the mode excitations arise from each wall decelerating and turning around at the intersection point. While this case is included here primarily for heuristic reasons, physically it corresponds to the case where a bubble nucleates instantaneously and expands out, and the result is consistent with particle production only at nucleation, with no particle production taking place during the expansion phase. The underlying physics of this result will be clarified in the final paragraph of this subsection (after case (d) below).  

\vskip 0.2cm
\noindent \textit{(d) Field excitation away from the minimum at collision point:}

The simple field ansatz used in the above examples, in \Eref{eq:cases} and \Eref{eq:newcase}, which match the ones used in the earlier studies \cite{Watkins:1991zt, Falkowski:2012fb}, miss one crucial ingredient. For both, the field value evaluates to $v_\phi$ at the origin. However, when the two bubbles walls completely superpose, the field value is known to add linearly, thus getting excited away from the true minima, jumping up to $2 v_\phi$ \cite{Jinno:2019bxw}. This piece is missing in the above treatments. We can incorporate this additional ingredient by adding to the field configuration a sharp jump at the origin, such as a sharp Gaussian of the form $e^{-(t^2+x^2)/\sigma}$, where $\sigma/v_\phi^2\gg 1$ determines how sharply the Gaussian is peaked around the origin. Since the Fourier transform is linear, the overall result can be obtained simply by calculating the Fourier transform of the above Gaussian and adding it to the previous results, e.g. \Eref{eq:FTelastic} and \Eref{eq:singlebubble}. Evaluating this numerically, we find that the Fourier transform of this Gaussian peak gives vanishing contributions to $\omega^2-k^2>0$ modes. This is also consistent with the numerical results from the realistic cases studied in \cite{Mansour:2023fwj} discussed below, which did consider field excitations to $2v_\phi$ at the collision point.

This result provides another insight into the underlying physics giving rise to the Fourier mode excitations. Field configurations that are symmetric in both space and time, such as the Gaussian peak above, cannot give rise to $\omega^2-k^2>0$ excitations. Exciting effectively massive Fourier modes with $\omega^2-k^2>0$ requires field dynamics that occurs faster in time than in space. From Fig.\,\ref{fig:simpleconfigurations}, we observe that this only occurs at the origin in panels (b) and (c). In panel (b), the field jumps from $0$ to $v_\phi$ and back to $0$ across the origin in the time direction (seen visually from the transition in color from blue to yellow to blue), but remains constant at $v_\phi$ in the spatial direction (yellow throughout). Likewise, in panel (c), we can see that the field changes from $0$ to $v_\phi$ in the time direction at the origin. At all other boundaries in the above simple configurations, the jump is symmetric, between $0$ and $v_\phi$ in both space and time. This is why only the actual collision process, i.e.\,the field dynamics at the collision point at the origin, creates $\omega^2> k^2$ excitations relevant for particle production.

\subsection{Realistic Cases}
\label{subsec:realistic}

The above simple cases capture the contributions from the collision between the bubble walls, but not from the scalar field excitations and oscillations that occur after the walls collide in any realistic setup. Here, we summarize the results from the literature for the efficiency factor $f(\chi)$ for realistic cases that incorporates these effects. 

Ref\,\cite{Falkowski:2012fb} presents analytic formulae in the idealized cases where the collision between two bubbles is either perfectly elastic or totally inelastic. For a perfectly elastic collision, where the walls bounce back perfectly with no energy dissipation in scalar oscillations, the efficiency factor is \cite{Falkowski:2012fb} 
\be
f_{\mathrm{PE}}(\chi)=\frac{16 v_{\phi}^2}{\chi^2}\, \text{Log}\left[\frac{2(\gamma_w/l_w)^2-\chi+2(\gamma_w/l_w)\sqrt{(\gamma_w/l_w)^2-\chi}}{\chi}\right]\,.
\label{eq:felastic}
\ee
For a totally inelastic collision, where the collision results in the bubble walls completely dissociating and all the energy going into scalar field oscillations, the factor is  \cite{Falkowski:2012fb} 
\begin{equation}
    f_{\mathrm{TI}}(\chi)= \frac{4 v_{\phi}^2 m_{t}^4 }{\chi^2 \left[ (\chi-m_{t}^2)^2+m_{t}^6\left(\frac{l_w}{\gamma_w}\right)^2 \right]} \mathrm{Log} \left[ \frac{2 \left(\frac{\gamma_w}{l_w}\right)^2 + \chi + 2\frac{\gamma_w}{l_w} \sqrt{\left(\frac{\gamma_w}{l_w}\right)^2+ \chi}}{\chi}\right]\,,
    \label{eq:fchiTI}
\end{equation}
where $m_t$ is the scalar mass in the true minimum.

While the above represent two idealized limits, a realistic collision consists of some energy going into scalar field oscillations in any collision, determined by the details of the scalar potential. Ref \cite{Mansour:2023fwj} studied such realistic collisions numerically, and found the following fit functions for the efficiency factor for elastic and inelastic collisions:
\be
f_{\text{elastic}}(\chi)= f_{\mathrm{PE}}(\chi)+\frac{v_{\phi}^2L_p^2}{15 m_{\mathrm{t}}^2}\exp{\left(\frac{-(\chi - m_{\mathrm{t}}^2+12 m_{\mathrm{t}}/L_p)^2}{440 \, m_{\mathrm{t}}^2 / L_p^2}\right)}\qquad \text{(elastic collisions)}
\label{eq:elasticfit}
\ee
\be
f_{\text{inelastic}}(\chi)= f_{\mathrm{PE}}(\chi)+\frac{v_{\phi}^2L_p^2}{4 m_{\mathrm{f}}^2}\exp{\left(\frac{-(\chi - m_{\mathrm{f}}^2+31 m_{\mathrm{f}}/L_p)^2}{650 \, m_{\mathrm{f}}^2 / L_p^2}\right)}\qquad \text{(inelastic collisions)}
\label{eq:inelasticfit}
\ee
Here, $f_{\mathrm{PE}}$ is the efficiency factor for a perfectly elastic collision as given in \Eref{eq:felastic}, and $m_t,\,m_f$ are the scalar masses in the true and false vacua respectively.  $L_p=\text{min}(R_*, \Gamma^{-1})$, where $\Gamma$ is the decay rate of the scalar and $R_*$ is the typical bubble size at collision, provides a measure of the extent to which scalar oscillations propagate in spacetime.

It is important to keep in mind that the above results are only applicable within a finite range of values of $\chi$, confined by physical scales in the setup. In the ultraviolet (UV), the results are not applicable for distances smaller than the boosted wall thickness $l_w/\gamma_w$, as effects related to the finite width of the bubble wall, which are not taken into account, become important. Likewise, the infrared (IR) cutoff is given by the inverse of the bubble wall radius $R_*^{-1}$; at physical scales larger than this, the existence of multiple bubbles should be taken into account. It is also worth mentioning that the above formulae ignore the bubble nucleation process. In practice, this is because the scale of bubble nucleation is significantly smaller than the length scale of the above studies, and cannot be properly resolved in the numerical approaches. While this shortcoming can be rectified, since the bubble nucleation phase is also amenable to the homogeneous approximation and analytic treatment, it is pragmatic to consider the nucleation phase separately (as done in Sec.\,\ref{subsec:nucleation}).

\subsection{Understanding Particle Production}

\begin{figure}
     \includegraphics[width=0.5 \textwidth]{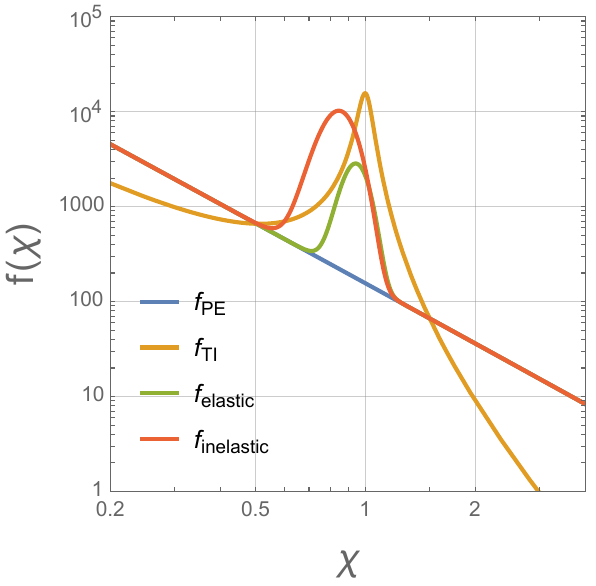} 
    \caption{Efficiency factor for particle production $f(\chi)$ in the four cases discussed in Sec.\,\ref{subsec:realistic}. For this plot, all masses are set to $v_\phi$, and we use $l_w=10/v_\phi, \,\gamma_w=200,$ and $L_p=200/v_\phi$. The axes ($f(\chi)$ and $\chi$ are plotted in appropriate units of $v_\phi$.}
    \label{fig:efffactors}
\end{figure}

The four functions for the efficiency factor from the previous subsection are plotted in \Fref{fig:efffactors} (the differences between the analytic formulae and numerical fits are discussed in detail in \cite{Mansour:2023fwj}). These results, along with the results for the Fourier transforms for the simple cases from Sec.\,\ref{sec:simple}, enable us to gain insight into the process of particle production. 

Note that both the analytic and numerical fit formulae consist of a combination of a power law and a peak in the spectrum. In all cases (except the totally inelastic case; the reason for this discrepancy is explained in \cite{Mansour:2023fwj}), the power law follows the form $f(\chi)\sim f_{PE}(\chi)\sim\chi^{-2}$ up to a logarithmic factor. This form of  $f_{\text{PE}}$ in the perfectly elastic collision limit follows from the Fourier transform in \Eref{eq:FTelastic}. As explained in the previous subsection, this component comes entirely from the collision between the bubble walls. This ubiquitous appearance of $f_{\mathrm{PE}}$, the efficiency factor for a perfectly elastic collision, can be understood from the discussions in the previous subsection. The first half of the collision, which consists of two bubbles slowing down as they overlap each other, is universal to all FOPTs, elastic or inelastic; this corresponds to the lower half of panel (b) in Fig.\,\ref{fig:simpleconfigurations}, and already contributes half of $f_{PE}$. The degree to which the collision is elastic or inelastic only becomes relevant in the second half of the collision process. Thus, the $f_{PE}$ component universally appears, within a factor of $2$, for all collisions.

The plots in \Fref{fig:efffactors} also show a peak in the spectrum, at the effective mass of the scalar field in the true (false) vacuum for inelastic (elastic) collisions. This component originates from oscillations of the background field around the relevant minimum after the field gets excited to unstable values as a result of the bubble collision. Crucially, note that this component is clustered around the scalar field mass, and does not contribute at higher $\chi$; therefore, this contribution is negligible for the production of particles significantly heavier than the scale of the phase transition.  

  \section{Particle Physics Aspects}
\label{sec:particlephysics}

The earlier sections have primarily focused on the spacetime configuration of the background field. This section is devoted to a discussion of some particle physics aspects relevant for particle production. 

\subsection{Non-universality of Particle Interactions and Masses} 
\label{subsec:nonuniversal}

During a FOPT, both the true and false vacua coexist simultaneously in different parts of space, and the Fourier transform gives the distribution of excitations over this mixed configuration. The nature of particle interactions, as well as particle masses, are generally different in the two vacua. Therefore, a crucial question arises: for the purposes of particle production -- in particular, the calculation of matrix elements in \Eref{optical} relevant for the decay of the excitations -- which vacuum is the correct one to use? Previous papers \cite{Watkins:1991zt,Konstandin:2011ds,Falkowski:2012fb} simply ignored this issue, choosing interactions and masses in the true vacuum; here we discuss some potential issues with this approach. 

For concreteness, consider again a scalar field $\psi$ that couples to the background  scalar field $\phi$ driving the phase transition via
   \be
 -\Delta \mathcal{L}_{\psi}= \frac{\lambda}{2} \phi^2 \psi^2 + \frac{1}{2} m_{0}^2 \psi^2,
            \label{eq:interaction}
            \ee
    where $\lambda$ is a dimensionless coupling and $m_{0}$ is a bare mass for $\psi$, which we include for generality. Again, let us take the scalar field vev to be $0$ and $v_\phi$ in the false and true vacua respectively. In the false vacuum, $\psi$ therefore only interacts with $\phi$ via a quartic term; consequently, an off-shell $\phi$ excitation can only decay into a three body final state $\phi^*\to \phi \psi\psi$. Note that the $\psi$ mass in the false vacuum is $m_0$.
    
In the true vacuum, one can also replace the $\phi$ field with its nonzero vev, so we have  
    \begin{equation}
           -\Delta \mathcal{L}_{\psi\,(true)}= \frac{\lambda}{2} \phi^2 \psi^2 + \lambda v_{\phi} \phi \psi^2 +  \frac{1}{2} (\lambda v_{\phi}^2 + m_{0}^2)\psi^2, 
            \label{eq.brokenint}
    \end{equation}
    Thus, in the true vacuum $\psi$ now interacts with $\phi$ via both a quartic and cubic term, so that the background scalar excitations can decay via $\phi^*\to\psi\psi$ in addition to the previously encountered $\phi^*\to \phi \psi\psi$.  The imaginary part of the 2-point 1PI Green's function for these two cases are given by 
\bea
  \mathrm{Im}\left( \Tilde{\Gamma}^{(2)}(\chi)\right)_{\phi^*\to\psi\psi} &=& \frac{\lambda^2 v_{\phi}^2}{8 \pi} \sqrt{1-\frac{4 m_{\psi}^2}{\chi}}\,   \Theta[\,\chi-(2 m_\psi)^2]\,,\nonumber\\
    \mathrm{Im}\left( \Tilde{\Gamma}^{(2)}(\chi)\right)_{\phi^*\to\phi\psi\psi} &=& \frac{\lambda^2 \chi}{512 \pi^3} \sqrt{1-\frac{(m_\phi+2 m_{\psi})^2}{\chi}}\,   \Theta[\,\chi-(m_\phi+2 m_{\psi})^2].
\label{eq:melements}
\eea
Moreover, in the true vacuum, the $\psi$ mass is $m_{\psi}^2=m_{0}^2+ \lambda v_{\phi}^2$. 
    
The $\psi$ interactions as well as mass are therefore vacuum dependent, and it becomes important to determine which vacuum is relevant for particle production\,\footnote{Note that this subtlety does not arise for the calculation of gravitational waves production, since the graviton coupling is universal.}. 

As discussed in the previous sections, particles get produced from the collision process (corresponding to the power law feature in the efficiency factor) as well as scalar waves -- oscillations of the scaler field around one of its minima (corresponding to the peak in the efficiency factor). For the latter contribution, it is clear that masses and interactions corresponding to the minima that the field is oscillating around --  true (false) vacuum for inelastic (elastic) collisions -- are the correct ones to use. The correct approach for the collision stage, on the other hand, is unclear, for several reasons.

In a bubble collision process, both the false and true vacua regions exist at the same time, on opposite sides of the bubbles. A particle with mass $m$ has a typical physical scale associated with it: its Compton wavelength $\sim m^{-1}$; as this is generally larger than the thickness of the boosted bubble walls at collisions, the physical length scale of the particle extends beyond the length scale of the collision process, and is sensitive to the existence of both vacua. Furthermore, the excitation of the massive modes of the background field that can decay into particles, as we saw earlier, comes from the bubble collision process, which corresponds to the background field jumping from the false to the true vacuum. The notion of a particle, on the other hand, is only well defined at a stable point of a theory, but not across a transient excitation between stable points. Moreover, since the bubble collision process involves the field jumping across two minima, it cannot be treated as a small perturbation away from either of the two vacua, further complicating the construction of a solution around a stable point of the theory.  

For particles for which both vacua are the same, i.e. particles with the same interactions and approximately the same mass in both vacua, these considerations are not problematic. This is true, for instance, for particles that are significantly heavier than the scale of symmetry breaking. On the other hand, for particles that are sensitive to this difference, e.g.\,for particles that obtain their mass primarily from the phase transition, this distinction becomes important. A rigorous resolution of this problem is lacking in the literature, and is beyond the scope of this work. Here we simply remark that since particles correspond to states that are long-lived compared to the collision process, it appears reasonable to use the properties corresponding to the vacuum state that is achieved after the collision process. Earlier papers on particles from bubble collisions \cite{Watkins:1991zt,Konstandin:2011ds,Falkowski:2012fb,Freese:2023fcr} did not discuss these caveats, but also followed this prescription.

\subsection{Non-perturbative Effects}
\label{subsec:resonant}

So far, we have not considered the possibility of nonperturbative, resonant enhancement of particle production. Ref.\,\cite{Watkins:1991zt} suggested that reaheating after FOPTs should occur in essentially the same way as after inflation. It is well known that nonperturbative processes such as tachyonic instability and parametric resonance can be important in such cases, resulting in explosive particle production, depending on the details of the scalar field dynamics  \cite{Traschen:1990sw,Shtanov:1994ce,Kofman:1997yn,Felder:2000hj,Felder:2001kt,Amin:2014eta,Shakya:2023zvs}. Furthermore, Ref.\,\cite{Zhang:2010qg} reported that parametric resonance can lead to very efficient particle production at bubble collisions. However, all of the standard studies of such resonant effects \cite{Traschen:1990sw,Shtanov:1994ce,Kofman:1997yn,Felder:2000hj,Felder:2001kt,Amin:2014eta,Shakya:2023zvs} are performed under the assumption that the background field evolution is spatially homogeneous. In particular, the study of bubble collisions in Ref.\,\cite{Zhang:2010qg} was also performed with this assumption, i.e.\,neglecting spatial gradient terms in the equation of motion. In this subsection, we qualitatively discuss the relevance of such resonant effects for particle production from FOPTs considering the highly inhomogeneous nature of the process. Here we will only discuss the case of tachyonic instability; the main arguments and conclusions apply analogously to the case of parametric resonance.  

Consider again the Klein Gordon equation for a scalar field $\psi$
\be
(\partial_t^2-\partial_x^2+m_{\text{eff},\psi}^2 (x,t))\,\psi(t,x)=0\,,
\label{kgetach}
\ee
where the effective mass $m_{\text{eff},\psi}$ can receive contributions from the background field undergoing the phase transition. In the homogeneous limit, one can decompose the $\phi$ field into modes of definite momenta. If  $m_{\text{eff},\psi}^2 (t) < 0 $, modes with momenta $k<|m_{\text{eff},\psi}|$ experience tachyonic instability and have an exponentially growing solution 
\be
    \psi_k(t)\propto\, \mathrm{exp}\,\left[\sqrt{m_{\text{eff},\psi}^2(t) -k^2}\,t\,\right]\,.
    \label{eq:tachyonic}
\ee
This results in a rapid coherent growth of particle number. 

When the setup is spatially inhomogeneous, the above result receives several modifications. First, the spatial derivative term in \Eref{kgetach} can become large, and it is well known that spatial derivatives have the effect of countering time derivatives, suppressing the rapid time evolution of the field. Furthermore, assuming the mass becomes tachyonic due to the effects of the background field dynamics, $m_{\text{eff},\psi}^2 (x,t) < 0 $ now only holds in finite regions of space, corresponding to the length scale of the boosted bubble walls or the scalar field oscillation scale $v_\phi\sim m_{t,f}$. As a result, \Eref{eq:tachyonic} is now only applicable at this, or smaller, length scales, i.e. for $k > v_\phi$. However, if the tachyonic mass is an outcome of the background field dynamics, it is plausible to assume $|m_{\text{eff},\psi}|<v_\phi$. Thus we see that the finite spatial extent of the dynamics tends to affect precisely the modes $k<|m_{\text{eff},\psi}|$ that could grow tachyonically. Moreover, since such particle production now only takes in a finite region of space, the produced particles no longer collect coherently at their production point but diffuse out in space. Such dissipation occurs over a timescale given by the scale of the inhomogeneity, $t\sim v_\phi^{-1}$. This timescale is shorter than the timescale over which the exponential growth of particle number occurs in \Eref{eq:tachyonic}, $t\sim m_{\text{eff},\psi}^{-1}$, which provides another argument for the suppression of tachyonic growth in such inhomogeneous configurations. 

Similar arguments apply to parametric resonance. The point, in essence, is that resonant processes such as tachyonic instability and parametric resonance require a $\textit{coherent}$ setting that seeds the growth of particle number. Such coherence is lost in FOPTs due to the inhomogeneous nature of the process: the mass term is only coherent over spatial distances $\sim v_\phi^{-1}$, and dissipation of particle number over space due to particle production in localized regions prevents a coherent growth of particle number beyond timescales $\sim v_\phi^{-1}$. Because of such factors, it is unlikely that such resonant processes can be very effective for particle production from FOPTs. It will be interesting to perform explicit (likely numerical) calculations incorporating such inhomogeneities to understand whether this is indeed the case. 

\section{Summary}
\label{sec:summary}

This paper discussed various aspects related to particle production from the dynamics of the background scalar field in a first order phase transition (FOPT), providing improved conceptual understanding of the underlying physics of the process beyond what currently exists in the literature. The main findings of this paper are summarized below:

\begin{itemize}

\item Nucleation stage: The process of bubble nucleation creates particles that couple to the background field. In the thin-wall approximation, this can be estimated using the same calculation as particle production out of vacuum in a homogeneous transition, and is given by \Eref{nucleationcontribution}. The number density of produced particles is parametrically $\sim l_w^{-3} (R_c/R_*)^3$. The physical scale of the process is $l_w^{-1}$, the inverse wall thickness, which also parameterizes the timescale of bubble nucleation; the $(R_c/R_*)^3$ factor accounts for diffusion of the particles from the bubble nucleation site to all space. This number density is significantly smaller than the number density of particles produced from bubble collisions, but could nevertheless be significant enough to backreact on and modify the nucleation process. 

\item Expansion stage: Bubble wall propagation at constant velocity does not lead to particle production (Sec.\,\ref{subsec:single}), whereas accelerating bubble walls, as realized in phase transitions with runaway behavior, are estimated to produce a particle number density $\sim R_*^{-3}$, where $R_*$ is the typical bubble size at collision (Sec.\,\ref{subsec:accelerating}). Since $R_*^{-1}$ is of the order of the Hubble scale, this contribution is also generally subdominant to the particle number density produced from the bubble collision process.

\item Bubble collision and post-collision stages: Particle production contributions from these stages are captured by \Eref{eq:ndensity},\,(\ref{number}), with the efficiency factor $f(\chi)$ given by \Eref{eq:elasticfit} for elastic collisions and \Eref{eq:inelasticfit} for inelastic collisions. $f(\chi)$ universally consists of a power law component $f(\chi)\sim \chi^{-2}$, which originates from the collision process, as deduced from analyzing the Fourier transforms of realistic collision configurations relative to a few simple cases in Sec.\,\ref{sec:expcol}, and a peak centered around the effective mass of the scalar, which comes from scalar field oscillations following bubble collisions.   

\item 
Particle interactions and masses are not universal, and take different forms in the two vacua, which coexist during a FOPT. The bubble collision process, in particular, involves a sharp, local transition of the background field from one vacuum to another, complicating the construction of a rigorous solution around a stable point of the theory. The resolution of this problem is unclear, but it appears reasonable to use the properties corresponding to the vacuum state that is realized after the collision process as an approximate solution.

\item Implications of the inhomogeneous nature of the process for nonperturbative, resonant effects were examined in Sec.\,\ref{subsec:resonant}. Since masses are not coherent over scales larger than $v_\phi^{-1}$, and particles diffuse out from the localized production points over a timescale $v_\phi^{-1}$, preventing a coherent buildup of particle number, such processes are likely suppressed and irrelevant for FOPTs. 

\end{itemize}

Overall, particle production from the dynamics of the background scalar field is found to be a significant aspects of first order phase transitions yet remains relatively understudied, especially compared to the more broadly and extensively studied phenomenon of gravitational waves production. Such particle production holds important implications for several relevant aspects: it can affect the dynamics of the background field, provide an additional source of gravitational waves related to FOPTs \cite{Inomata:2024rkt}, as well as address outstanding problems in particle physics and cosmology, such as dark matter and baryogenesis (as explored in \cite{Falkowski:2012fb,Katz:2016adq,Freese:2023fcr,Giudice:2024tcp,Cataldi:2024pgt}), and therefore remains a crucial ingredient of FOPTs that merits further attention. In this context, the improved understanding of the particle production process at various stages of a FOPT provided by this paper invites further investigation into various related matters.  In particular, several results here, such as particle production from accelerating bubbles, the choice of the parameters in the case of non-universal masses and interactions, as well as the arguments for the inefficiency of nonperturbative processes, are based on simple intuitive arguments, but could (and should) be further sharpened with more rigorous calculations.

\acknowledgments
The author is grateful to Gian Giudice, Thomas Konstandin, Hyun Min Lee, Henda Mansour, Alex Pomarol, and Geraldine Servant for several detailed discussions on various aspects of this work. The author is supported by the Deutsche Forschungsgemeinschaft under Germany’s Excellence Strategy - EXC 2121 Quantum Universe - 390833306.

\appendix

\section{Particle Production from a Single Propagating Bubble Wall}
\label{appendixsinglewall}

The goal of this appendix is to understand particle production using the standard Bogoliubov transformation approach for a scalar $\psi$ that follows the Klein Gordon equation from Eq.\,\ref{kgew}:
\be
\left(-\frac{\partial^2}{\partial t^2}+\frac{\partial^2}{\partial x^2}\right)\psi(t,x)=\frac{m_\psi^2}{4}\left(1-\text{tanh}\left(\frac{v_w t-x}{l_w}\right)\right)^2\psi(t,x)\,.
\label{kgeapp}
\ee
If the field $\psi$ is in its vacuum state in the false vacuum, it can be written as a collection of harmonic oscillators with negative frequencies $\sim\int \frac{dp}{\sqrt{2E}}\,e^{-i\omega t+ipx}$, where $\omega$ is determined via the standard dispersion relation $w^2=p^2+m^2$. 
Following the standard approach of Bogoliubov transformations, our goal is to expand the late time asymptotic solution $\psi(vt>x)$ in terms of its mode functions $\psi_p$, expressing the mode functions in the form
\be
\psi_{p}\sim \frac{1}{\sqrt{2E}}\,(\alpha_p \,e^{-i\omega t} +\beta_p\,  e^{i\omega t})
\ee
The positive frequency modes can be interpreted as particles, with the number density of particles produced (with momentum p) given by the square of the Bogoliubov coefficient in the above expansion, $n_p=|\beta_p|^2.$

To obtain this solution, note that the mass term in  Eq.\,\ref{kgeapp} depends on both variables $t$ and $x$. Hence we first change variables to eliminate this double dependence by switching to the new coordinates $(x_1,x_2)$ given by
\be
x_1=\frac{1}{\sqrt{1+v_w^2}} (v_w t-x),~~~x_2=\frac{1}{\sqrt{1+v_w^2}} (t+v_w x)\,. \label{newcoordinates}
\ee
With these new coordinates, we can use the relation 
\be
g'_{\mu v_{\phi}}(x')=\frac{\partial x^\rho}{\partial x'^\mu}\frac{\partial x^\sigma}{\partial x'^v_{\phi}} g_{\rho\sigma}(x)
\ee
to rewrite the general KG equation
\be
(g^{\alpha\beta}\partial_\alpha\partial_\beta-m^2)\psi=0
\ee
as
\be
\left[\,\left(\frac{1-v_w^2}{1+v_w^2}\right)\left(-\frac{\partial^2}{\partial x_2^2}+\frac{\partial^2}{\partial x_1^2}\right)-\frac{4v_w}{1+v_w^2}\frac{\partial}{\partial x_1}\frac{\partial}{\partial x_2}\right]\psi(x_1,x_2)={m_\psi^2(x_1)}\psi(x_1,x_2)\,,
\label{kge2}
\ee
where the mass term now only varies along $x_1$:
\be
m_\psi^2(x_1)=\frac{m_\psi^2}{4}\left(1-\text{tanh}\left(\frac{\sqrt{1+v_w^2}\,x_1}{l_w}\right)\right)^2\,.
\ee
The physics is now invariant under translations along $x_2$, hence its conjugate momentum $p_2$ is a conserved quantity. We can therefore expand the field $\psi$ in terms of its $p_2$ modes, which are plane waves, and solve the modified KG equation Eq.\,\ref{kge2} mode-by-mode (note that such momentum decomposition is only justified if the corresponding momentum is conserved, hence this choice of coordinates) to construct our solution. For each $p_2$ mode, Eq.\,\ref{kge2} gives the following equation for its coefficient, which we write as $\psi_{p2}(x_1)$:
\be
\left(\frac{\partial^2}{\partial x_1^2}-\frac{4iv_w p_2}{1-v_w^2}\frac{\partial}{\partial x_1}\right)\psi_{p2}(x_1)=\left[\frac{1+v_w^2}{1-v_w^2}{m_\psi(x_1)^2}-p_2^2\right]\psi_{p2}(x_1)\,.
\label{KGnew}
\ee
We anticipate being able to write the asymptotic form of the solution in terms of a superposition of plane waves $\sim e^{ip_1x_1+ip_2x_2}$. Recall that the initial vacuum state configuration is a superposition of negative frequency modes $\sim\int \frac{dp}{\sqrt{2E}}\,e^{-i\omega t+ipx}$. Rewriting these in terms of the new coordinates, we get
\be
e^{-i\omega t+ipx}=\text{exp}\left[-\frac{\omega v_w+p}{\sqrt{1+v_w^2}}x_1-\frac{\omega-pv_w}{\sqrt{1+v_w^2}}x_2\right] \,.
\ee
The conjugate of $x_2$, which corresponds to a conserved quantity, is 
\be
p_2=-\frac{\omega-pv_w}{\sqrt{1+v_w^2}}=-\frac{|p|-pv_w}{\sqrt{1+v_w^2}}\,, 
\label{p2form}
\ee
where we have used $\omega=|p|$ for the initial state since $\psi$ is initially massless. Note that there is a degeneracy: two different values of p (one positive, one negative) map on to the same value of $p_2$. The conjugate of the other coordinate $x_1$ is given by
\be
p_1=-\frac{\omega v_w+p}{\sqrt{1+v_w^2}}\,.
\ee
This quantity is not conserved but changes across across the bubble wall. 

Next, we examine the dispersion relation in our new coordinates. For a plane wave $\sim e^{ip_1x_1+ip_2x_2}$, Eq.\,\ref{kge2} enforces the following dispersion relation between the conjugate momenta:
\be
p_1^2-p_2^2-\frac{4v_w}{1-v_w^2}p_1p_2+m_\psi^2\frac{1+v_w^2}{1-v_w^2}=0.
\ee
This can be rewritten as a solution for $p_1$ in terms of $p_2$:
\be
p_1=\frac{2v_wp_2}{1-v_w^2}\pm\sqrt{\frac{p_2^2(1+v_w^2)^2-m_\psi^2(1-v_w^4)}{(1-v_w^2)^2}}\,.
\label{p1p2}
\ee
This represents the mixing between modes due to the change in mass across the bubble wall. The two values correspond to the transmitted and reflected parts of the plane wave of conjugate momentum mode $p_2$ when it encounters the bubble wall, and the final asymptotic form can be written as a superposition of these two parts for each mode.  

Next, we rewrite the general asymptotic form of the plane wave in the new coordinates in terms of the physical space and time coordinates $x,t$:
\be
e^{ip_1x_1+ip_2x_2}\sim \text{exp}\left[\frac{p_1v_w+p_2}{\sqrt{1+v_w^2}}t+\frac{p_2v_w+p_1}{\sqrt{1+v_w^2}}x\right]\,.
\ee
We are interested in the coefficient of the time coordinate in the exponential, $p_1v_w+p_2$, at late times. In particular, since particles correspond to positive frequencies, we are interested in scenarios where this coefficient can be positive, i.e.
\be
\frac{p_1v_w+p_2}{\sqrt{1+v_w^2}}>0\,.
\ee 
Substituting in $p_1$ from Eq.\,\ref{p1p2}, this condition can be written as 
\be
p_2\,\frac{1+v_w^2}{1-v_w^2}\pm v_w\sqrt\frac{p_2^2(1+v_w^2)-m_\psi^2(1-v_w^4)}{(1-v_w^2)^2}>0\,.
\label{condition}
\ee
The two values, again, correspond to the frequencies of the transmitted and reflected parts of the incoming plane wave. 

From Eq.\,\ref{p2form}, we have $p_2=-({|p|-pv_w})/{\sqrt{1+v_w^2}}$. If $p>0$, we have $p_2=-p(1-v_w)/\sqrt{1+v_w^2}<0$ for $v_w<1$. Likewise, if $p<0$, then $p_2=p(1+v_w)/\sqrt{1+v_w^2}<0$ again. Thus, we conclude that the first term in Eq.\,\ref{condition} is always negative for $v_w<1$ for $p_2$ given by Eq.\,\ref{p2form}, i.e.\,any component of the initial vacuum (negative frequency) state.  Thus, the condition in Eq.\,\ref{condition} can only be satisfied if the second term (featuring the square root) is bigger than the first term, and for the positive sign. The condition for this, after a bit of algebra, is 
\be
p_2^2(v_w^4-1)>m_\psi^2(1-v_w^2)\,.
\ee
Note that the above calculation and result is independent of the exact form of the mass term in Eq.\,\ref{kgeapp}, and remains applicable as long as the change of coordinates restricts the mass variation to a single coordinate.  

\bibliography{references}

\end{document}